# Could Cirrus Clouds Have Warmed Early Mars?

Ramses M. Ramirez[i,ii], James F. Kasting[iii,iv,v]
[i] Carl Sagan Institute, Cornell University, Ithaca, NY, 14853, USA
[ii] Cornell Center for Astrophysics and Planetary Science, Cornell University, Ithaca, NY, 14853, USA
[iii] Department of Geosciences, Penn State University, University, Park, PA,16802, USA
[iv] Penn State Astrobiology Research Center, Penn State University University Park, PA,16802, USA
[v] NASA Astrobiology Institute Virtual Planetary Laboratory, University of Washington, 98195, USA

Corresponding author: Ramses Ramirez
Cornell University
304 Space Sciences Building
Ithaca, NY, USA 14850
email: rramirez@astro.cornell.edu
ph: 1(480)-296-6477

Keywords: Mars, atmosphere; Mars, climate; Atmospheres, evolution

**ABSTRACT**

The presence of the ancient valley networks on Mars indicates that the climate at 3.8 Ga was warm enough to allow substantial liquid water to flow on the martian surface for extended periods of time. However, the mechanism for producing this warming continues to be debated. One hypothesis is that Mars could have been kept warm by global cirrus cloud decks in a $CO_2$-$H_2O$ atmosphere containing at least 0.25 bar of $CO_2$ (Urata and Toon, 2013). Initial warming from some other process, e.g., impacts, would be required to make this model work. Those results were generated using the CAM 3-D global climate model. Here, we use a single-column radiative-convective climate model to further investigate the cirrus cloud warming hypothesis. Our calculations indicate that cirrus cloud decks could have produced global mean surface temperatures above freezing, but only if cirrus cloud cover approaches ~75 - 100% and if other cloud properties (e.g., height, optical depth, particle size) are chosen favorably. However, at more realistic cirrus cloud fractions, or if cloud parameters are not optimal, cirrus clouds do not provide the necessary warming, suggesting that other greenhouse mechanisms are needed.

# INTRODUCTION

The climate of early Mars continues to be an issue of major astrobiological interest. Nearly all investigators recognize that the formation of large-scale features, specifically ancient valleys like Nanedi Vallis or Warrego Vallis, required a flowing liquid, almost certainly water, These ancient valleys are found almost exclusively on heavily-cratered terrain, indicating that they formed at or before ~ 3.8–3.6 Ga (Fassett and Head, 2011). However, debate has persisted as to how warm a climate was required, as well as how long such a warm period could have lasted. Some authors have argued that dense $CO_2$ and $H_2O$ atmospheres could have warmed early Mars above the freezing point of water, producing the required runoff rates to form the ancient valleys (Pollack et al., 1987; Forget and Pierrehumbert, 1997). But the combination of increased Rayleigh scattering at elevated pressures and $CO_2$ condensation limit modeled mean surface temperatures for such $CO_2$-$H_2O$ rich atmospheres to ~220 - 230 K (Kasting, 1991; Tian et al., 2010; Ramirez et al., 2014a; Forget et al., 2013; Wordsworth et al., 2013). Forget and Pierrehumbert (1997) showed that infrared backscatter from $CO_2$ ice clouds can provide the required warming as long as fractional cloud cover approaches 100%. However, warming from this mechanism is substantially reduced if more realistic cloud cover fractions (~50%) are assumed (Mischna et al., 2000; Forget et al., 2013). Moreover, Kitzmann et al. (2013) showed that when the 2-stream radiative transfer algorithm used in previous studies is replaced with a more accurate discrete ordinates method, this mechanism is even less effective.

The difficulty in producing warm climates using $CO_2$-$H_2O$ atmospheres (e.g. von Paris et al., 2015) has prompted solutions in which the greenhouse effect is supplemented by other gases, such as $SO_2$ and $CH_4$ (Johnson et al., 2008; Ramirez et al., 2014a; Halevy and Head, 2014; Kerber et al., 2015). But $SO_2$ forms sulfate aerosols which act to cool the climate (Tian et al., 2010; Kerber et al. 2015), whereas strong near-infrared absorption by $CH_4$ causes stratospheric warming at the expense of surface warming (Ramirez et al., 2014a, Fig. S10). One greenhouse gas combination does work, however. Ramirez et al. (2014a) demonstrated that a dense $CO_2$ atmosphere containing ~5–20% $H_2$ could have raised mean surface temperatures above the freezing point of water, perhaps explaining the ancient valley networks. More recently, photochemical calculations by Batalha et al. (2015) showed that such $H_2$ concentrations could have been maintained if volcanic outgassing rates were a few times higher than that on modern Earth, or if hydrogen escaped from early Mars more slowly than the diffusion limit.

Other authors have proposed various "cold early Mars" models to explain these features. Of these, the impact hypothesis (Segura et al., 2002, 2008, 2012) has been the most widely cited. According to this idea, large impacts during the Late Heavy Bombardment period could have generated transient warm climates with erosion rates high enough to produce the valley networks. In these calculations, these authors estimated the amount of water that could have been vaporized both from large impacts and from ground ice mobilized by the collision. This water was then assumed to have rained out and carved the valleys. Segura et al. (2008) predicted that 650 m of global rainfall could have been generated by this process and that the martian climate could have been intermittently warm for hundreds to thousands of years. In contrast, later authors have argued that much more water over much longer timescales was needed to form the valleys (e.g. Barnhardt et al., 2009; Hoke et al., 2011). For example, Hoke et al., (2011) find that intermittent runoff averaging ~ 10 cm/yr for 30 – 40 million years was needed to form the larger valley networks. This requires ~ $10^4$ times more rainfall than predicted by the impact hypothesis.

In the most recent iteration of this hypothesis, Segura et al. (2012) have advocated that very large asteroid and comet impacts would have raised surface temperatures hundreds of degrees, triggering a runaway greenhouse state. The authors suggest that this runaway state may have been long-lasting, greatly amplifying the amount of rainfall that could have been generated. For this to be true, a runaway greenhouse state must be sustainable (in a cloud-free atmosphere) at an absorbed solar flux of ~80 W/m$^2$, according to their Fig. 5. This requires that the emitted thermal-infrared flux from a cloud-free runaway greenhouse must be equal to that same value. But other recent authors who have calculated such fluxes (e.g., Pierrehumbert, 2010, Fig. 4.37: Kopparapu et al., 2013; Goldblatt et al., 2013, Ramirez et al., 2014b) all find outgoing IR fluxes of ~280 W/m$^2$, which is a factor of 3.5 higher than the Segura et al. value. So, a cloud-free steam atmosphere should not have been self-sustaining on early Mars.

There may still be a way to make the impact hypothesis work, however, if clouds can stabilize a warm climate. Using a 3-D climate model, Urata and Toon (2013) argued that cirrus clouds could have kept early Mars warm for long time periods if the cloud particles were 10 μm or larger and if cloudy grid cells were fully cloud covered. In contrast, none of the simulations with partially-covered cloudy grid cells yielded warm solutions. An initially warm state, perhaps caused by an impact, is required. Their results differ from 3-D GCM results by Wordsworth et al. (2013), who found very little warming from water clouds (including cirrus clouds), even assuming fully cloud-covered grid cells. Both of these studies were concerned with how to best treat cloud overlap in 3-D models to better simulate those clouds that are smaller than the model grid resolution.

Another important parameter in these models is the precipitation efficiency of cloud particles, which directly impacts cloud lifetimes. In the simulations just discussed, Wordsworth et al. (2013) employed a precipitation threshold that initiated above a certain cloud mass. However, when this precipitation threshold was disabled, Wordsworth et al. (2013) found that $H_2O$ clouds became much thicker optically, leading to significant surface warming (up to 250 – 260 K) before a slow decline to cooler temperatures (200 – 215 K). The mean atmospheric densities of condensed $H_2O$ in these simulations were unrealistically high compared to estimated values for the Earth under snowball conditions (Abbot et al., 2012), leading those authors to conclude that these latter simulations were not physically plausible.

By comparison, Urata and Toon (2013a) parameterized the conversion of cirrus cloud particles to snow using a modified version of the method of Lin (1983), which is based on terrestrial observations. The Lin model contains an auto-conversion factor ($B$) that controls the rate at which cloud particles are converted to snow. Urata and Toon decreased this auto-conversion rate to as low as 1% of the rates computed in the Lin (1983) formulation. Warm climates were obtained only by invoking such decreases. The authors justified this assumption by arguing that martian clouds should form at higher altitudes than on Earth, increasing the time required for the particles to fall out. We return to this question in the Discussion section, where we argue that such long residence times for cloud particles are physically unrealistic.

In addition, other aspects of this mechanism have yet to be investigated, so we do so here. This is the first study (to our knowledge) to report on the effectiveness of cirrus clouds on early Mars as a function of fractional cloud cover. Moreover, these previous studies have not reported the effect that cirrus cloud thicknesses opacities have on the effectiveness of this mechanism. Both of these questions can be most readily addressed with our single-column radiative-convective climate model because of its relatively fine wavelength resolution (93 total intervals in the solar and infrared), as well as its speed, which allows us to quickly compare and contrast different scenarios. Here, we use our single-column model and utilize scaling relations to analyze the cirrus cloud mechanism of Urata and Toon (2013) for different crystal sizes, cloud thicknesses, and cloud fractions to determine whether the cirrus-warming mechanism is feasible.

## 2. METHODS

*Climate model description*

We used a single-column radiative-convective model first developed by Kasting et al. (1984) and updated most recently by Ramirez et al. (2014a, 2014b). In recent versions of this model (Kopparapu et al., 2013; Ramirez et al., 2014a), the atmosphere is divided into 100 logarithmically spaced layers that extend from the ground to a specified low pressure at the top of the atmosphere ($\sim 10^{-4}$ bar). For this study, we increased the number of layers to 200 to improve the vertical resolution of thin cirrus cloud decks (see next subsection). Absorbed and emitted radiative fluxes in the stratosphere are assumed to be balanced. Should the radiative lapse rate exceed the moist adiabatic value at lower atmospheric layers, the model relaxes to a moist $H_2O$ adiabat at high temperatures, or to a moist $CO_2$ adiabat when it is cold enough for $CO_2$ to condense (Kasting, 1991). This defines a convective troposphere which we assume to be fully saturated. This last assumption overestimates the actual greenhouse effect, but it avoids the problem of specifying an arbitrary vertical distribution of relative humidity. We can tolerate an error in this direction, as our calculations suggest that surface warming is limited even in this fully saturated case.

As usual in 1-D climate modeling, we assumed that the planet is flat and the Sun remains fixed at a zenith angle of 60º. A modern average martian solar flux of 585 W/m$^2$ was assumed in most of the calculations. As in previous studies by our group (Kasting, 1991; Ramirez et al., 2014a), we adjusted the surface albedo of our model to a value (0.216) that yields the mean surface temperature for present day Mars (~ 218 K). Although Goldblatt and Zahnle (2011) have argued that this methodology may overestimate the greenhouse effect of dense early atmospheres, we consider this error to be acceptable because, even with this assumption, it is difficult to make early Mars warm using cirrus clouds. By contrast, our neglect of $CO_2$ clouds may cause us to underestimate the greenhouse effect (Forget and Pierrehumbert, 1997; Mischna et al., 2000). However, this error is also acceptable because the warming expected from $CO_2$ clouds in dense $CO_2$ atmospheres is no more than a few degrees (Kitzmann et al., 2013; Forget et al., 2013).

A standard two-stream approximation is employed for both the solar and thermal-infrared portions of the radiative code (Toon et al., 1989). Although this scheme exhibits substantial errors for some scattering phase functions (Kitzmann et al., 2013), it remains appropriate for $H_2O$ clouds. Our correlated-$k$ coefficients parameterize gaseous absorption across 38 solar spectral intervals spanning from 0.2 to 4.5 microns (~2000 – 50,000 cm$^{-1}$) and 55 thermal-infrared intervals ranging from 0 – 15,000 cm$^{-1}$ (> ~ 0.66 microns). We used separate 8-term coefficients for both $CO_2$ and $H_2O$. These coefficients were calculated using line width truncations of 500 cm$^{-1}$ and 25 cm$^{-1}$, respectively, and were computed over 8 temperatures (100, 150, 200, 250, 300, 350, 400, 600 K), and 8 pressures (10$^{-5}$ – 100 bar) (Kopparapu et al., 2013; Ramirez et al. 2014ab). Overlap between gases was computed by convolving the $k$-coefficients for the two different greenhouse gases within each broadband spectral interval.

Far-wing absorption in the 15-micron band of $CO_2$ is parameterized by utilizing the 4.3-micron $CO_2$ region as a proxy (Perrin and Hartmann, 1989). In analogous fashion, the BPS water continuum of Paynter and Ramaswamy (2011) is overlain over its entire spectral range of validity (0 - ~18,000 cm$^{-1}$). Collision-induced absorption (CIA) between $CO_2$ molecules is parameterized using a relatively recent formulation (Gruszka and Borysow, 1998; Baranov et al., 2004). Finally, we used measured water vapor Rayleigh scattering coefficients for use in standard parameterizations (Vardavas and Carver, 1984) utilizing available data (Bucholtz, 1995; Edlén, 1966). We also incorporated data for Rayleigh scattering by $CO_2$ (Allen, 1976).

*Cirrus cloud parameterization*

We investigated the cloud feedback of Urata and Toon (2013a) using a methodology similar to that of Kasting (1988) and Ramirez (2014b), except that we expect ice clouds, not liquid water clouds, at the mostly sub-freezing temperatures exhibited here. As cirrus clouds on early Mars form predominantly in the $H_2O$ moist convective region (Wordsworth et al., 2013), we assume that the highest level at which these cloud decks can be placed is the boundary between the $CO_2$ and $H_2O$ moist convective regions. Unless stated otherwise, our baseline simulations assumed thin (~ 1 km thick) cirrus cloud decks, so as to maximize cirrus cloud warming. Fractional cloud cover is modeled by averaging cloudy and cloud-free radiative fluxes (e.g. Arking et al., 1996). More specifically, the fluxes at 50% cloud cover were determined by averaging the respective clear-sky and 100% cloud cover values for $F_{IR}$ and $F_S$ (parameters defined in the next section). In analogous fashion, $F_{IR}$ and $F_S$ at 75% cloud cover were computed by averaging the corresponding fluxes at 50% and 100% cloud cover.

Because we are interested in comparing our results to those of Urata and Toon (2013a), we used the same Mie optical properties for the particles, which they were kind enough to provide in digital form. Following Urata and Toon (2013a) and Ramirez (2014b), the wavenumber-dependent optical depths ($\tau$) were obtained from the following expression:

$$\tau = \frac{3 \cdot Q_{eff} \cdot IWC \cdot \Delta z}{4 \cdot r \cdot \rho} \qquad (1)$$

Here, $Q_{eff}$ is the wavenumber-dependent extinction efficiency, $\rho$ is the mass density of ice, $r$ is the effective particle radius, $IWC$ is the ice water content (g/m$^3$), and $\Delta z$ is the vertical path length of the layer (m). The $IWC$ was computed by assuming that it scales linearly with local pressure ($P$), following Kasting (1988). According to Platt et al. (1997), in Earth's atmosphere $IWC/P \sim 0.33$ at $P = 0.6$ bar. Although vertical wind speeds within cirrus clouds decks for early Mars conditions are unknown, recent 3-D simulations for a 0.5 bar atmosphere suggest that upper-tropospheric wind speeds ranging from ~ 0 to 10s m/s are not unreasonable (Fig. 8-9 in Wordsworth et al., 2013). These values are similar to the corresponding vertical wind speeds measured over Cape Canaveral or Christmas Island (Endlich and Singleton, 1969; Gage et al., 1991). Thus, if we assume that the vertical wind velocities for Earth and Mars are roughly similar (following Kasting, 1988), given that the force that suspends the cloud particles is proportional to the product of vertical wind speed and atmospheric density (ibid), this ratio should then be multiplied by the relative ratio of planetary gravities (1/0.38), yielding $IWC/P \sim 0.88$ for cirrus clouds on Mars. We assume that this same ratio of $IWC/P$ applies at all heights in the atmosphere. We also performed sensitivity studies to evaluate the effect of changing this assumption.

*Climate modeling procedures*

To gauge the radiative forcing of cirrus clouds on early Mars, we adopted an isothermal stratosphere with a temperature of 155 K. Then, we assumed a mean surface temperature of 273 K (corresponding to a warm enough climate that generates the water amounts necessary to form the ancient valleys, e.g. Hoke et al., 2011) and integrated a moist adiabat upwards until it intersected the stratospheric temperature profile (see, e.g., Kasting, 1988; Ramirez et al., 2014b). Clouds were then placed at various heights within the convective (adiabatic) troposphere. This methodology is sometimes termed 'inverse modeling'. Using these assumed temperature profiles and cloud decks, we computed the outgoing infrared flux, $F_{IR}$, and the absorbed solar flux, $F_S$, for the present martian solar constant. Then we computed the *effective* solar flux, $S_{eff} = F_{IR}/F_S$, needed to sustain that surface temperature, following the methodology of Kasting (1988, 1991). $S_{eff}$ thus represents the brightness of the early Sun, compared to today, that would have been needed to maintain a mean surface temperature at the freezing point of water. For cloudy atmospheres in which this latter condition is satisfied, we then checked our calculations by increasing the mean surface temperature until radiative-convective equilibrium is achieved, assuming early Mars insolation (solar flux = 75% that of present).

## 3. RESULTS

*Studies involving 1-km thick cirrus clouds*

To test whether cirrus clouds could have warmed early Mars, we studied fully saturated $CO_2$ atmospheres with pressures ranging from 0.5 bar (the value assumed by Urata and Toon (2013)) to 3 bar (Fig. 1). Our single-column climate model assumes that the convective troposphere can be modeled as a region in which the lapse rate follows an $H_2O$ moist adiabat underlies a colder region in which the lapse rate follows the $CO_2$ moist adiabat (Fig. 1). Cirrus cloud decks were placed at different heights only within the $H_2O$ moist adiabatic region. This assumption is in agreement with 3-D simulations by Wordsworth et al. (2013) which demonstrate that virtually all water clouds should be located in the $H_2O$ moist adiabatic region. Parameters are listed in Table I.

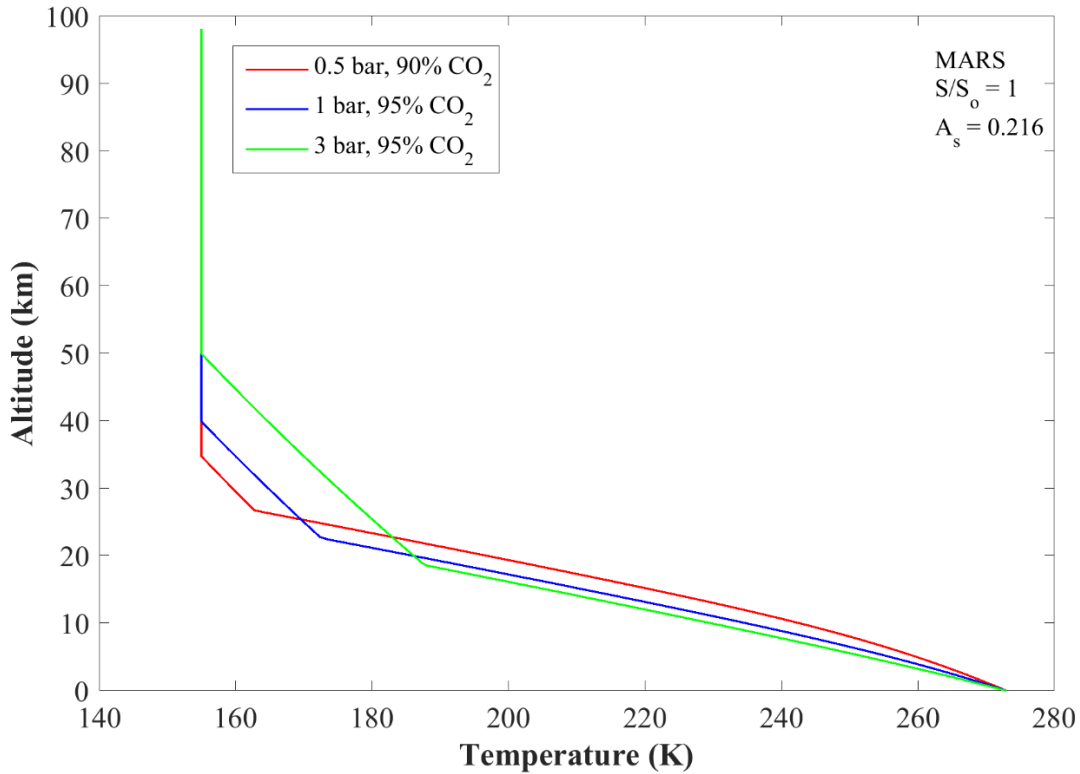

**Fig. 1:** Assumed temperature profiles for three different atmospheres: 1) 0.5 bar surface pressure, 90% $CO_2$ (red), 2) 1 bar surface pressure, 95% $CO_2$ (blue), and 3) 3 bar surface pressure, 95% $CO_2$. The remaining background gas is $N_2$. The constant temperature line is the stratospheric region, with temperature fixed at 155 K. This is underlain by the $CO_2$ moist convective region, which is underlain by the liquid water moist adiabat region that extends down to the surface. The mean surface temperature is set to 273 K for all three cases.

**Table Ia.** Associated properties for 1-km cloud decks composed of 10 µm cirrus particles for the 0.5 bar $CO_2$ atmosphere [a]

| P | ALT | T (K) | FIR | FS | SEFF | PALB | IWC (g/m$^3$) | Tau(0.55 microns) | Tau(10 microns) |
|---|---|---|---|---|---|---|---|---|---|
| 0.492 | 0.402 | 272.1 | 137.3 | 49.29 | 2.79 | 0.6628 | 0.433 | 66.1 | 77.33 |
| 0.384 | 3.88 | 263.1 | 133.2 | 47.23 | 2.82 | 0.677 | 0.338 | 50.7 | 59.24 |
| 0.273 | 8.44 | 248.5 | 122.5 | 43.83 | 2.79 | 0.7001 | 0.24 | 40.1 | 46.89 |
| 0.207 | 11.95 | 234.7 | 107.3 | 43.62 | 2.46 | 0.7016 | 0.182 | 31.79 | 37.19 |
| 0.157 | 15.21 | 220.0 | 89.74 | 45.59 | 1.97 | 0.6882 | 0.138 | 24.67 | 28.85 |
| 0.113 | 18.86 | 202.5 | 69.27 | 50.38 | 1.375 | 0.6554 | 9.94(-2) | 17.77 | 20.79 |
| 7.88(-2) | 22.49 | 184.3 | 51.13 | 57.29 | 0.892 | 0.6081 | 6.94(-2) | 12.24 | 14.31 |
| 5.57(-2) | 25.69 | 168.1 | 38.46 | 64.79 | 0.594 | 0.5569 | 4.9(-2) | 8.42 | 9.846 |
| 5(-5) | 82.64 | 155 | 138.7 | 108.8 | 1.275 | 0.256 | - | - | - |

**Table Ib**. Associated properties for 1-km cloud decks composed of 10 µm cirrus particles for the 1 bar $CO_2$ atmosphere [a]

| P | ALT | T (K) | FIR | FS | SEFF | PALB | IWC (g/m$^3$) | Tau(0.55 microns) | Tau(10 microns) |
|---|---|---|---|---|---|---|---|---|---|
| 0.976 | 0.424 | 271.7 | 110.4 | 45.75 | 2.41 | 0.6871 | 0.859 | 137.81 | 161.19 |
| 0.748 | 4.06 | 259.5 | 106.9 | 41.16 | 2.6 | 0.7184 | 0.658 | 102.57 | 119.96 |
| 0.518 | 8.76 | 240.42 | 97.75 | 35.96 | 2.72 | 0.754 | 0.456 | 77.68 | 90.85 |
| 0.384 | 12.31 | 224.09 | 84.84 | 35.11 | 2.42 | 0.76 | 0.339 | 59.57 | 69.68 |
| 0.287 | 15.57 | 208.27 | 70.39 | 36.58 | 1.92 | 0.7498 | 0.252 | 44.9 | 52.52 |
| 0.20 | 19.21 | 190.08 | 54.32 | 40.4 | 1.34 | 0.7273 | 0.177 | 31.3 | 36.61 |
| 0.154 | 21.72 | 177.18 | 44.4 | 44.25 | 1.0 | 0.6973 | 0.1352 | 23.63 | 27.64 |
| 5(-5) | 86.39 | 155 | 111 | 102.1 | 1.09 | 0.3018 | - | - | - |

**Table Ic**. Associated properties for 1-km cloud decks composed of 10 µm cirrus particles for the 3 bar $CO_2$ atmosphere [a]

| P | ALT | T (K) | FIR | FS | SEFF | PALB | IWC (g/m$^3$) | Tau(0.55 microns) | Tau(10 microns) |
|---|---|---|---|---|---|---|---|---|---|
| 2.91 | 0.469 | 271.19 | 75.73 | 44.1 | 1.717 | 0.6984 | 2.558 | 453.58 | 530.52 |
| 2.16 | 4.46 | 254.64 | 75.3 | 37.91 | 1.985 | 0.7407 | 1.903 | 323.37 | 378.11 |
| 1.44 | 9.53 | 231.43 | 72.51 | 31.57 | 2.297 | 0.784 | 1.267 | 230.7 | 269.84 |
| 1.034 | 13.3 | 213.22 | 65.9 | 29.5 | 2.234 | 0.7982 | 0.91 | 169.25 | 197.96 |
| 0.691 | 17.5 | 192.33 | 55.03 | 29.61 | 1.859 | 0.7974 | 0.608 | 113.05 | 132.22 |
| 5(-5) | 96.43 | 155 | 75.79 | 87.87 | 0.8625 | 0.3990 | - | - | - |

[a]Values are listed at the midpoint of the cloud deck. Read 1.0(-5) as 1x10$^{-5}$.

Panel b in Fig. 2 shows the effective solar flux, $S_{eff}$ (=$F_{IR}/F_S$), needed to maintain the surface temperature in steady state. Here, $F_{IR}$ and $F_S$ are the net outgoing infrared and net absorbed solar fluxes, respectively, at the top of the atmosphere. Values of $S_{eff} > 1$ indicate that a surface temperature of 273 K could only be maintained in Mars' future, when the Sun has brightened relative to today. The value of $S_{eff}$ at 3.8 Ga, when most of the martian valleys are thought to have formed, was ~0.75 (Gough, 1981). This value is shown by a dashed horizontal line in Fig. 2. As one can see from this figure, values of $S_{eff} < 0.75$ can only be found for the 0.5 bar scenario and only for an extremely limited range of cloud heights. Remember also that these calculations assume 100 percent cloud cover. If fractional cloud cover is below ~75 percent, no solutions are found with $S_{eff} < 0.75$ (see discussion below).

Another way of analyzing these results, following Goldblatt et al. (2013), is to assess the effect of clouds on the difference between $F_{IR}$ and $F_S$. In panel c of Figure 2, the incident solar flux has been normalized differently, so that $F_{IR} - F_S = 0$ when clouds are not present. (This requires a higher incident solar flux because these fluxes do not balance at $Seff = 0.75$). Under these assumptions, values of $F_{IR} - F_S > 0$ indicate that clouds cool the climate, while negative values suggest that they warm it. Warming by clouds is possible over a greater range of cloud heights, although as we have seen, this does not mean that it would have been sufficient to warm early Mars. The results of our cloud study, using 10 micron-sized particles, are summarized in Fig. 2 and Table I. The efficacy of cirrus cloud warming increases with decreasing atmospheric pressure. 1-km thick clouds produce the greatest surface warming when they are centered at ~0.056 bar, 0.15 bar, and ~0.7 bar for the 0.5, 1, and 3 bar atmospheres, respectively (Table I), (Cirrus clouds located at pressures lower than ~0.7 bar for the 3-bar atmosphere are unphysical because these locations are above the moist convective region). Furthermore, Fig. 2c shows that the maximum amount of cirrus cloud warming occurs in the 0.5 bar atmosphere, with lesser warming at 1 and 3 bar surface pressures. This is because the $CO_2$ condensation region is smallest in the 0.5 bar atmosphere, moving the top of the moist convective region, and thus the cirrus cloud deck, to higher altitudes where cloud warming is enhanced (Figure 1). Panels b and c in subsequent figures are graphed using the same approach as that just described for Fig. 2.

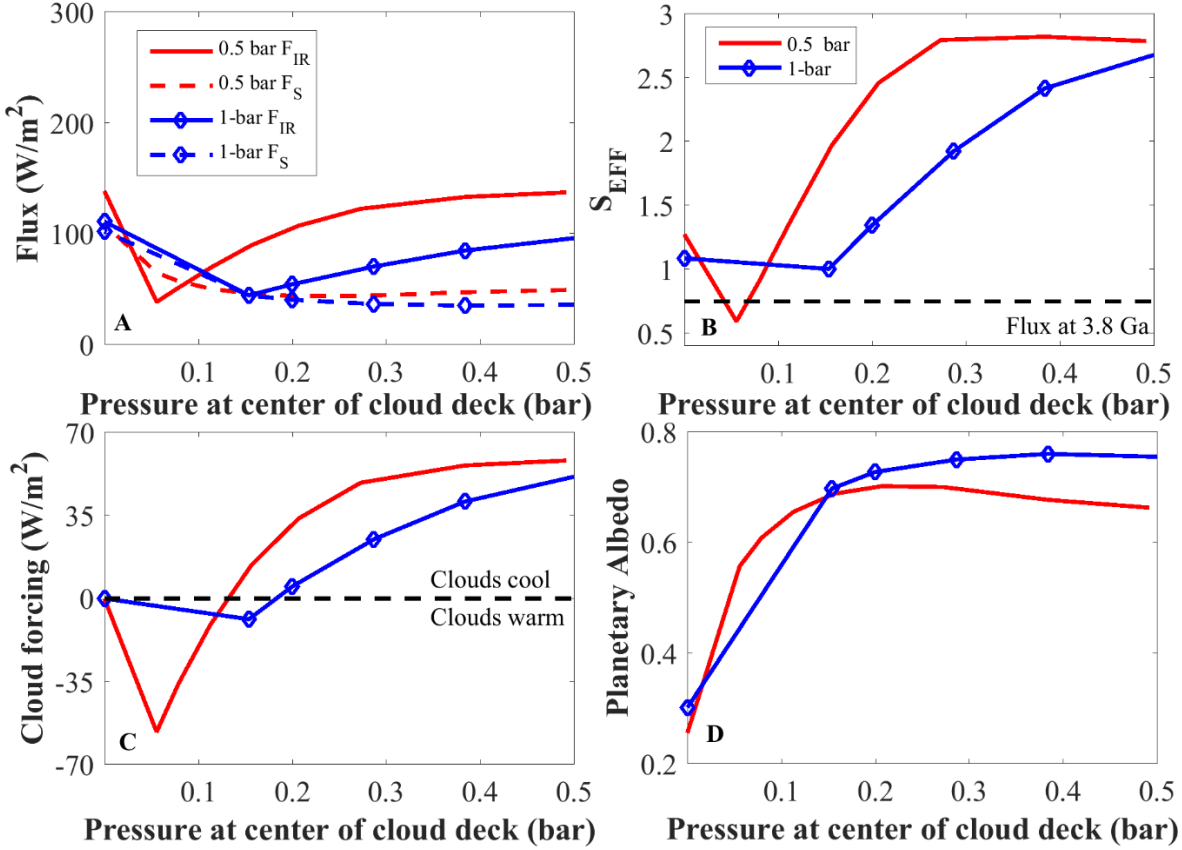

**Fig. 2:** Effect of a single cirrus cloud layer on various model parameters for the 0.5 (red) and 1-bar (blue open diamond) fully-saturated, $CO_2$ atmospheres of Fig. 1 using 10-micron cloud particles. In Panels A and B, the assumed solar flux is that at present Mars. Panel C is normalized differently (see text). Panel A: net absorbed solar flux, $F_s$, (dashed) and outgoing infrared flux, $F_{IR}$ (solid), at the top of the atmosphere; Panel B: effective solar flux, $S_{eff}$; Panel C: net flux change, $F_{IR} - F_S$; Panel D: planetary albedo. The horizontal axis shows the pressure at the center of the assumed 1-km-thick cirrus cloud deck. The 3-bar atmosphere from Fig.1 is not shown because corresponding cloud decks are at higher pressures than illustrated.

To determine the effect of other assumptions made in our model, we performed additional cloud calculations for the 0.5-bar surface pressure case, placing clouds at the pressure level in which the 1-km thick cirrus clouds are most effective (0.0557 bar). First, we analyzed the effect of fractional cloud cover (50, 75%, 100%) over a range of IWC spanning from 1/3 to 10 times the baseline value, which is $4.9 \times 10^{-2}$ g/m$^3$ at this pressure level (Table Ia). Higher IWC values represent wetter atmospheric conditions than predicted by Clausius-Clapeyron, simulating large influxes of water vapor, such as those resulting from giant impacts (i.e. Urata and Toon, 2013).

Results are shown in Fig. 3 and Table II. As expected, the $S_{eff}$ values required to warm early Mars are lowest at complete cloud coverage for all IWC values. Decreased reflectivity below 100% cloud cover results in lower planetary albedos (Figure 3d). For complete cloud coverage, the baseline (x1) and x1/3 IWC cases suggest that early Mars could be warmed to the freezing point at $S_{eff}$ values less than ~ 0.6 (Table IIa). Thus, as $S_{eff}$ was equal to ~0.75 at 3.8 Ga, there should have been plenty of solar energy to warm the red planet if the clouds were actually this thick and if fractional cloud cover was this high.

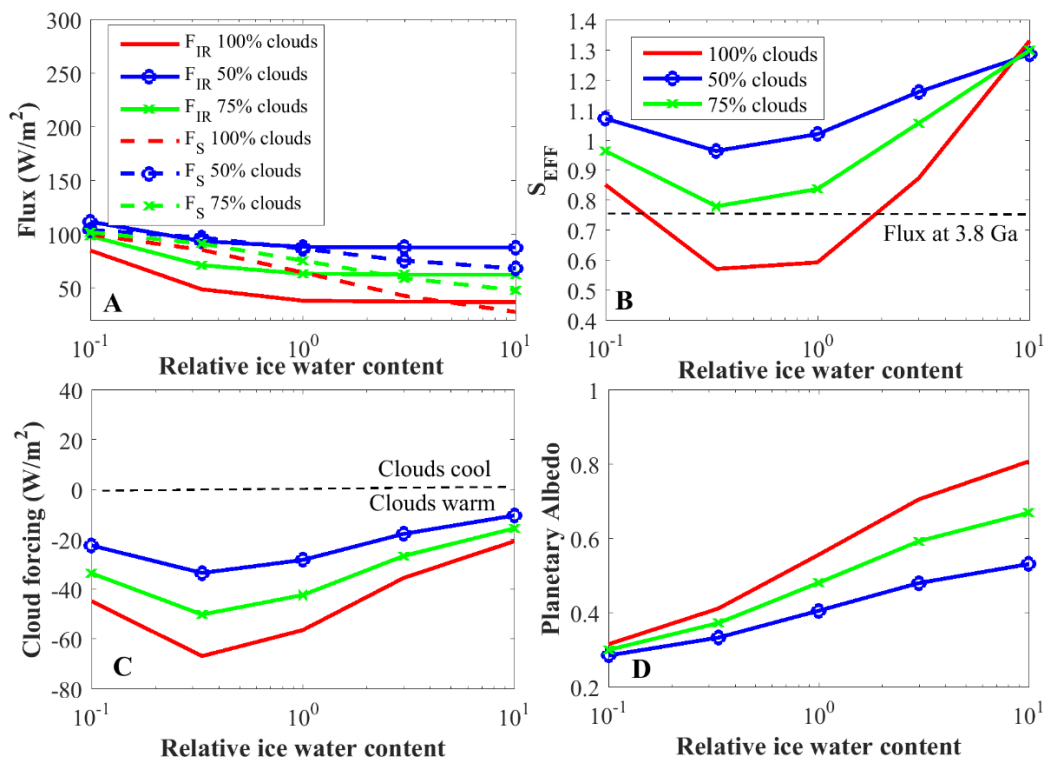

**Fig. 3:** Effect of a single cirrus cloud layer on the 0.5 bar atmosphere from Fig. 1 using 1-km thick clouds composed of 10-micron cloud particles. In Panels A and B, the assumed solar flux is that at present Mars. Panel C is normalized differently (see text). The horizontal axis shows the cloud water ice content relative to that in the Fig. 2 case for 100% (red), 50% (blue open circle), and 75% (green with cross) cloud cover fractions. Panel A: net absorbed flux, $F_s$ (dashed) and outgoing flux, $F_{IR}$ (solid) at the top of the atmosphere. Panel B: effective solar flux, $S_{eff}$; Panel C: net flux change, $F_{IR} - F_S$; Panel D: planetary albedo. As before, $S_{eff}$ values below 0.75 (dashed horizontal line in panel B) indicate cloud forcing sufficient to maintain above-freezing mean surface temperatures.

**Table IIa: Sensitivity study for a 0.5 bar $CO_2$ present Mars atmosphere with 1-km thick cloud decks using 10-micron cloud particles at P = $5.57 \times 10^{-2}$ bar with 100 % cloud cover**

| IWC  | FIR   | FS    | SEFF  | PALB   | TAU(0.55 microns) | TAU(10 microns) |
|------|-------|-------|-------|--------|-------------------|-----------------|
| x1   | 38.46 | 64.79 | 0.594 | 0.5569 | 8.419             | 9.847           |
| x3   | 37.58 | 42.99 | 0.874 | 0.706  | 25.26             | 29.54           |
| x10  | 37.39 | 28.07 | 1.332 | 0.8080 | 84.19             | 98.47           |
| x1/3 | 49.14 | 85.93 | 0.572 | 0.4123 | 2.78              | 3.25            |
| x1/10| 85.27 | 99.99 | 0.853 | 0.3161 | 0.842             | 0.985           |

**Table IIb: Sensitivity study for a 0.5 bar $CO_2$ present Mars atmosphere with 1-km thick cloud decks using 10-micron cloud particles at P = $5.57 \times 10^{-2}$ bar with 75 % cloud cover**

| IWC   | FIR75 | FS75  | SEFF75 | PALB   |
|-------|-------|-------|--------|--------|
| x1    | 63.54 | 75.79 | 0.8383 | 0.4816 |
| x3    | 62.88 | 59.44 | 1.0577 | 0.5934 |
| x10   | 62.73 | 48.25 | 1.3001 | 0.6699 |
| x1/3  | 71.55 | 91.65 | 0.7807 | 0.3731 |
| x1/10 | 98.64 | 102.2 | 0.9653 | 0.3010 |

**Table IIc: Sensitivity study for a 0.5 bar $CO_2$ present Mars atmosphere with 1-km thick cloud decks using 10-micron cloud particles at P = $5.57 \times 10^{-2}$ bar with 50 % cloud cover**

| IWC   | FIR50  | FS50   | SEFF50 | PALB   |
|-------|--------|--------|--------|--------|
| x1    | 88.61  | 86.8   | 1.021  | 0.4063 |
| x3    | 88.17  | 75.90  | 1.1617 | 0.4808 |
| x10   | 88.08  | 68.44  | 1.287  | 0.5318 |
| x1/3  | 93.95  | 97.37  | 0.9649 | 0.3339 |
| x1/10 | 112.02 | 104.40 | 1.073  | 0.2858 |

In analogous fashion, we also varied fractional cloud cover and IWC for the same 0.5 bar atmosphere but used larger (100-micron) particles (Fig. 4 and Tables III-IV). Again, for all IWC values, $S_{eff}$ values required to warm early Mars are lowest at 100% cloud cover. However, planetary albedo values are always lower for the 100-micron case. This is because larger particles at a given cloud mass translate to fewer particles with larger surface area, and so less energy is reflected to space.

Cloud greenhouse warming is even more effective with the larger particle size. In these simulations, early Mars could be warmed at $S_{eff}$ values approaching 0.5 (Figure 4; Table IVa), although this warming occurs at higher (x3 and x10, respectively) IWC values. At larger particle sizes, cloud opacities decrease according to eqn.1, requiring a higher IWC to generate the same amount of warming. Thus, our model, like that of Urata and Toon (2013), predicts that cirrus clouds produce more warming at 100-micron mean particle radius than they do at 10-micron radius, ignoring other factors such as sedimentation rates (see Discussion). However, we find that at 75% cirrus cloud cover, 3x IWC provides just enough warming to bring early Mars to the freezing point (Figure 4; Table IVb). In contrast, 75% cirrus cloud cover using 10-micron particles does not produce this amount of warming (Figure 3; Table IIb).

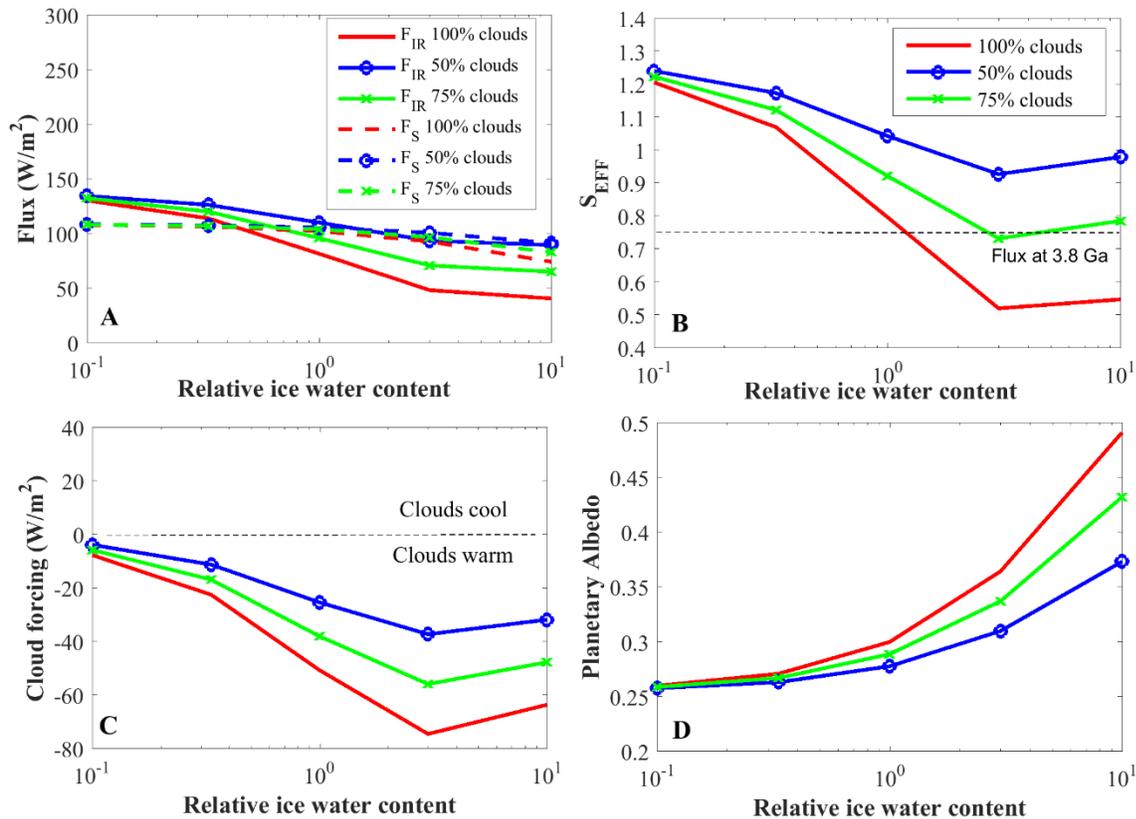

**Fig. 4:** Same as as in Fig.3 but using 1-km thick cirrus clouds composed of 100-micron cloud particles in the 0.5 bar atmosphere from Fig. 1

**Table III.** Associated properties for 1-km thick cloud decks composed of 100 µm cirrus particles for the 0.5 bar $CO_2$ atmosphere [a]

| P | ALT | T (K) | FIR | FS | SEFF | PALB | IWC (g/m$^3$) | Tau(0.55 microns) | Tau(10 microns) |
|---|---|---|---|---|---|---|---|---|---|
| 0.492 | 0.402 | 272.1 | 137.9 | 86.12 | 1.6 | 0.4109 | 0.433 | 6.61 | 7 |
| 0.384 | 3.88 | 263.1 | 134.0 | 87.96 | 1.524 | 0.3984 | 0.338 | 5.07 | 5.37 |
| 0.273 | 8.44 | 248.5 | 124.8 | 89.05 | 1.4 | 0.391 | 0.24 | 4.25 | 4.01 |
| 0.207 | 11.95 | 234.7 | 113.3 | 91.1 | 1.24 | 0.377 | 0.182 | 3.179 | 3.369 |
| 0.157 | 15.21 | 220.0 | 99.78 | 93.67 | 1.065 | 0.3593 | 0.138 | 2.47 | 2.614 |
| 0.113 | 18.86 | 202.5 | 85.91 | 96.92 | 0.886 | 0.3371 | 9.94(-2) | 1.77 | 1.88 |
| 7.88(-2) | 22.49 | 184.4 | 79.33 | 100.0 | 0.793 | 0.3156 | 6.94(-2) | 1.22 | 1.296 |
| 5.57(-2) | 25.7 | 168.1 | 81.76 | 102.5 | 0.798 | 0.299 | 4.9(-2) | 0.842 | 0.892 |
| 5(-5) | 82.64 | 155 | 138.7 | 108.8 | 1.275 | 0.257 | - | - | - |

[a]Values are listed at the midpoint of the cloud deck. Read 1.0(-5) as 1x10$^{-5}$.

**Table IVa:** Sensitivity study for a 0.5 bar $CO_2$ present Mars atmosphere composed of 1-km thick clouds composed of 100-micron particles at P = 5.57x10$^{-2}$ bar with 100 % cloud cover

| IWC | FIR | FS | SEFF | PALB | TAU(0.55 microns) | TAU(10 microns) |
|---|---|---|---|---|---|---|
| x1 | 81.76 | 102.49 | 0.798 | 0.30 | 0.842 | 0.892 |
| x3 | 48.36 | 92.9 | 0.521 | 0.3645 | 2.53 | 2.68 |
| x10 | 40.75 | 74.4 | 0.548 | 0.4913 | 8.42 | 8.92 |
| x1/3 | 114.1 | 106.59 | 1.206 | 0.271 | 0.281 | 0.294 |
| x1/10 | 130.4 | 108.13 | 1.071 | 0.2611 | 8.42(-2) | 8.92(-2) |

**Table IVb:** Sensitivity study for a 0.5 bar $CO_2$ present Mars atmosphere composed of 1-km thick clouds composed of 100-micron cloud particles at P = 5.57x10$^{-2}$ bar with 75 % cloud cover

| IWC | FIR75 | FS75 | SEFF75 | PALB |
|---|---|---|---|---|
| x1 | 95.99 | 104.07 | 0.922 | 0.2889 |
| x3 | 70.94 | 96.88 | 0.732 | 0.3373 |
| x10 | 65.25 | 83 | 0.786 | 0.4324 |
| x1/3 | 120.27 | 107.14 | 1.123 | 0.2672 |
| x1/10 | 132.5 | 108.3 | 1.224 | 0.2589 |

**Table IVc: Sensitivity study for a 0.5 bar $CO_2$ present Mars atmosphere composed of 1-km thick clouds composed of 100-micron cloud particles at $P = 5.57 \times 10^{-2}$ bar with 50 % cloud cover**

| IWC | FIR50 | FS50 | SEFF50 | PALB |
|---|---|---|---|---|
| x1 | 110.22 | 105.65 | 1.043 | 0.2778 |
| x3 | 93.52 | 100.85 | 0.9273 | 0.31 |
| x10 | 89.71 | 91.6 | 0.979 | 0.3735 |
| x1/3 | 126.39 | 107.7 | 1.174 | 0.2633 |
| x1/10 | 134.6 | 108.47 | 1.241 | 0.2578 |

*Sensitivity studies with thicker clouds and mean surface temperatures*

In the real world cloud thicknesses vary greatly. Although we cannot self-consistently compute cloud thicknesses in a 1-D model, we can prescribe different cloud thickness scenarios and assess their radiative effects. Here, we investigate the magnitude of greenhouse warming of thicker 5- and 10-km cirrus clouds using 10- and 100-micron particles for the baseline 0.5 bar atmosphere of Fig 1. For the 5-km clouds IWC is $6.19 \times 10^{-2}$ g/m$^3$ (at $P = 7.03 \times 10^{-2}$ bar), whereas for the 10-km clouds it is $8.04 \times 10^{-2}$ g/m$^3$ (at $P = 9.13 \times 10^{-2}$ bar) (Tables VI-VII).

Results are similar to those in the previous section, although thicker clouds generally produce less greenhouse warming than do thinner clouds (Figures 5–12; Tables V-VII). This is because, at low optical depths, the greenhouse effect of high-altitude cirrus clouds tends to dominate the increase in planetary albedo (e.g., Choi and Ho, 2006; Ramirez et al., 2014b). Once the clouds become too thick, the albedo increase outstrips the warming effect (Figures 5–12; Tables V–VII). Thus, our thin cirrus cloud simulations described above should provide an upper bound for the greenhouse warming from this mechanism.

As with both our own 1-km thick cirrus cloud simulations and those of Urata and Toon (2013), however, this mechanism is more effective for 100-micron cloud particles than it is for 10-micron particles (Figures 5–12). Our 5-km cirrus clouds composed of 100-micron particles can support a warm early Mars atmosphere at 1xIWC with an $S_{eff}$ of ~0.55 (Figures 7 -8; Table VIb). In contrast, a somewhat higher $S_{eff}$ (~0.6) at 1/10 IWC is required for 10-micron cloud particles (Figure 6; Table Va). No warm solutions are obtained with 10-km thick cirrus clouds composed of 10-micron particles (Figures 9-10). However, two warm solutions are found at 100% cloud cover if the clouds are composed of 100-micron particles and an $S_{eff}$ of ~0.65–0.7 is required at 1/3 and 1x IWC, respectively (Figures 11 – 12). All of these thick cloud warm solutions required 100% cloud cover.

Finally, we computed mean surface temperatures for the warm solutions for our 0.5 bar atmosphere at the different IWC values (Table VIII). Mean surface temperatures ranged from ~ 273 K to 285 K, exactly as would be expected from the inverse calculations described above.

*Comparison with Urata and Toon (2013)*

We compared the cloud forcing values of our 0.5 bar 10-micron cloud particle atmosphere versus the equivalent Case 17 of Urata and Toon (2013). According to Fig. 16 of their paper, cloud forcing for their warm, full-grid cloud cover case ranges from 53 W/m$^2$ to 30 W/m$^2$, averaging ~43 W/m$^2$ for the year. These cloud forcing values are most similar to those obtained by our 100-micron particle simulations at 50% and 75% cloud cover (Figures 4, 7-8, 11-12). Our model obtains a maximum cloud forcing of ~74 W/m$^2$ assuming 100% cloud cover of 1-km thick cirrus clouds composed of 100-micron particles at 3x IWC (Figure 4). This maximum forcing is higher than that of Urata and Toon (2013) because cloud coverage in their simulations was always less than 100% (Brian Toon, personal communication). The maximum cloud forcing decreases to ~65 W/m$^2$ at 1/3 IWC for clouds composed of 10-micron particles (Figure 3). Cloud forcing decreases dramatically for thicker clouds. With 100 micron-particles, the cloud forcing ranges from ~20 – 65 W/m$^2$ for 5-km and 20 – 60 W/m$^2$ 10-km thick clouds, respectively (Figure 8 and Figure 12). At the smaller particle size, this decreases to ~10 - 60 W/m$^2$ for 5-km thick clouds (Figure 6) and ~10 – 40 W/m$^2$ for 10-km thick clouds (Figure 10).

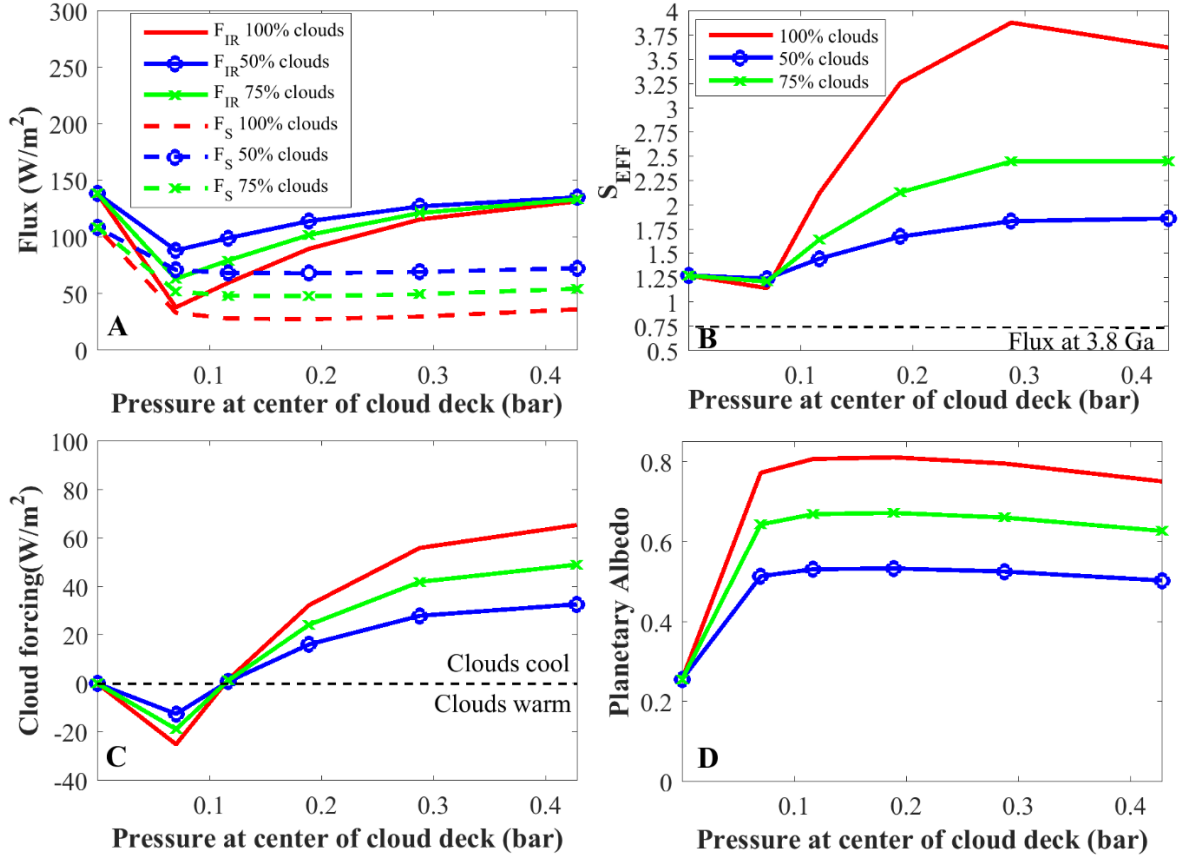

**Fig. 5:** Effect of a single cirrus cloud layer on various model parameters for the 0.5 bar fully-saturated $CO_2$ atmosphere of Fig. 1 using 5-km cirrus cloud decks composed of 10-micron particles for 100% (red), 50% (blue open circle), and 75% (green cross) cloud cover fractions. In Panels A and B, the assumed solar flux is that at present Mars. Panel C is normalized differently (see text). Panel A: net absorbed solar flux, $F_s$, (dashed) and outgoing infrared flux, $F_{IR}$ (solid), at the top of the atmosphere; Panel B: effective solar flux, $S_{eff}$; Panel C: net flux change, $F_{IR} - F_S$; Panel D: planetary albedo. The horizontal axis shows the pressure at the center of the assumed 1-km-thick cirrus cloud deck.

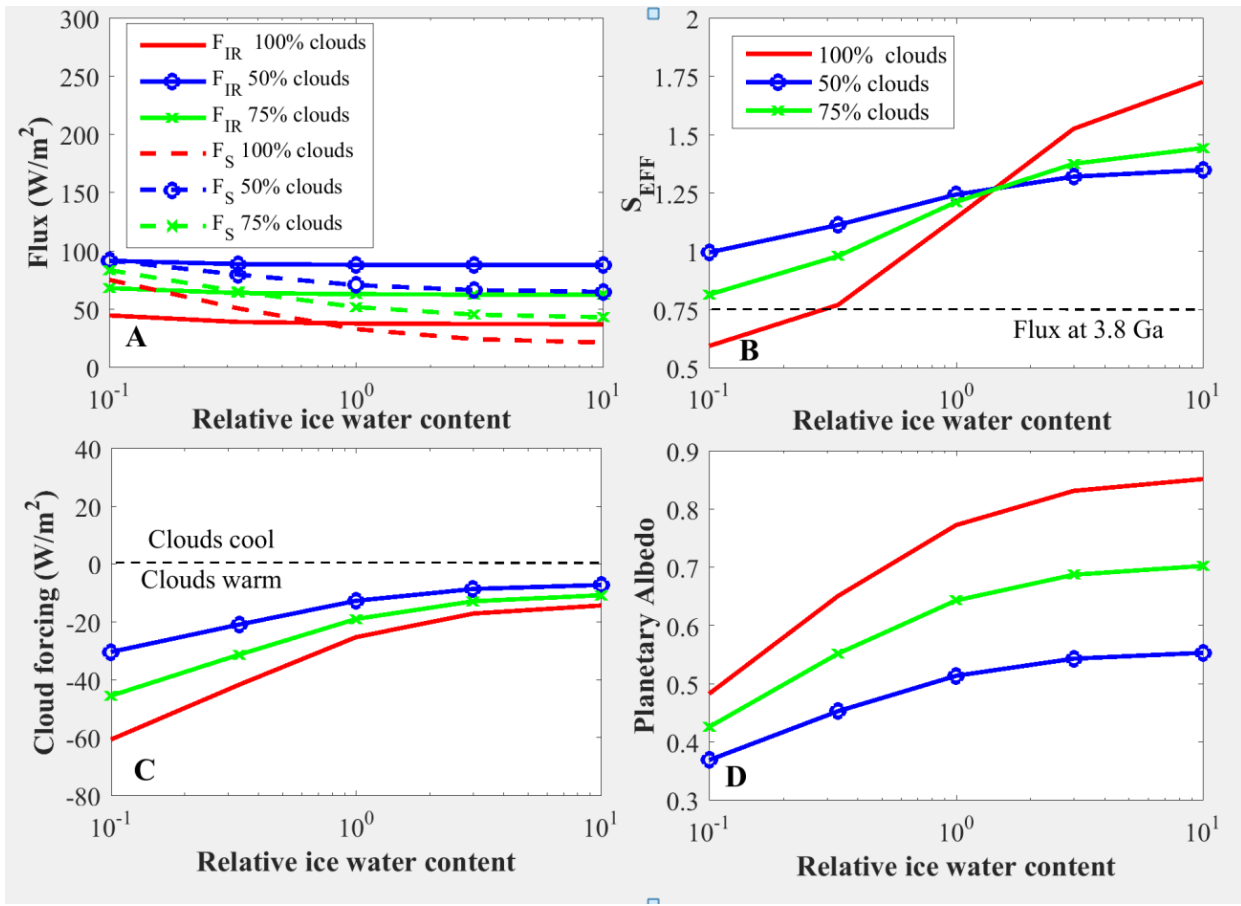

**Fig. 6:** Same as in Fig.3 but using 5-km thick cirrus clouds composed of 10-micron cloud particles in the 0.5 bar atmosphere from Fig. 1

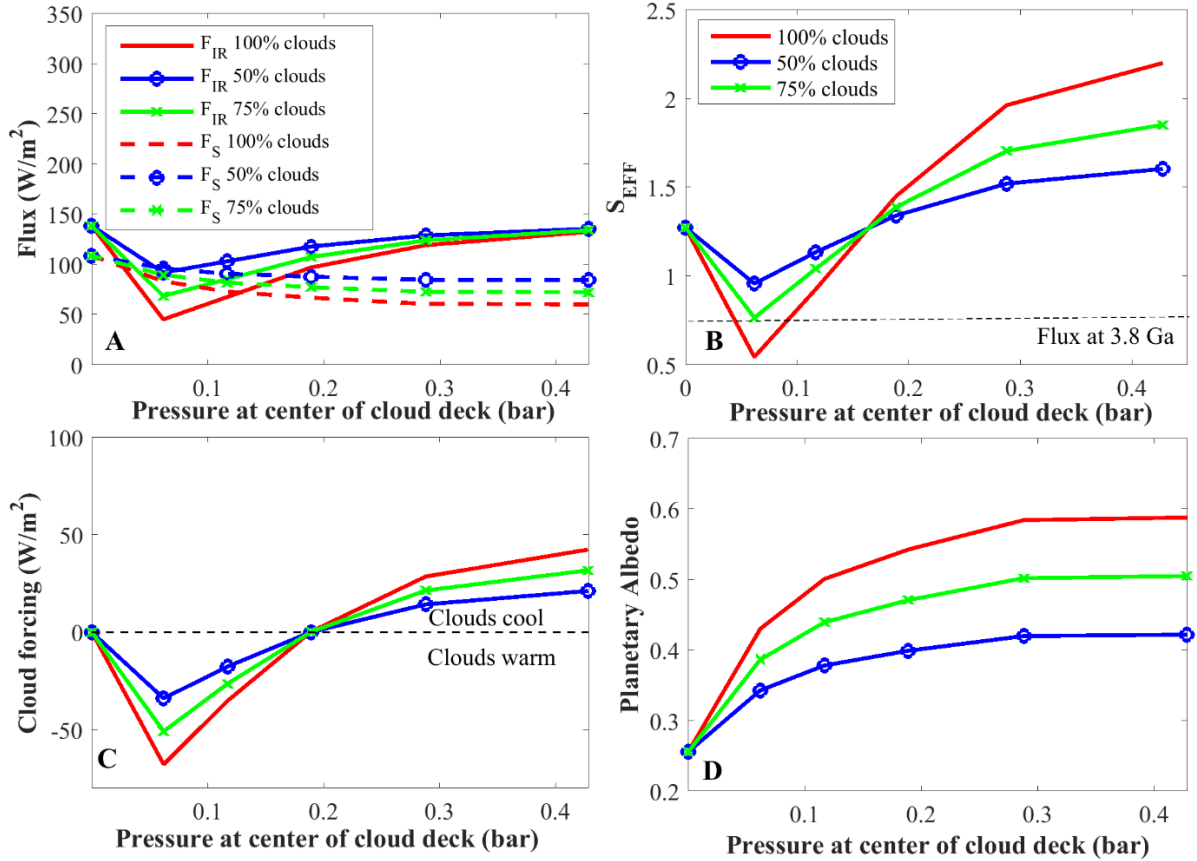

**Fig. 7:** Same as in Fig.5 but using 5-km thick cirrus clouds composed of 100-micron cloud particles in the 0.5 bar atmosphere from Fig. 1

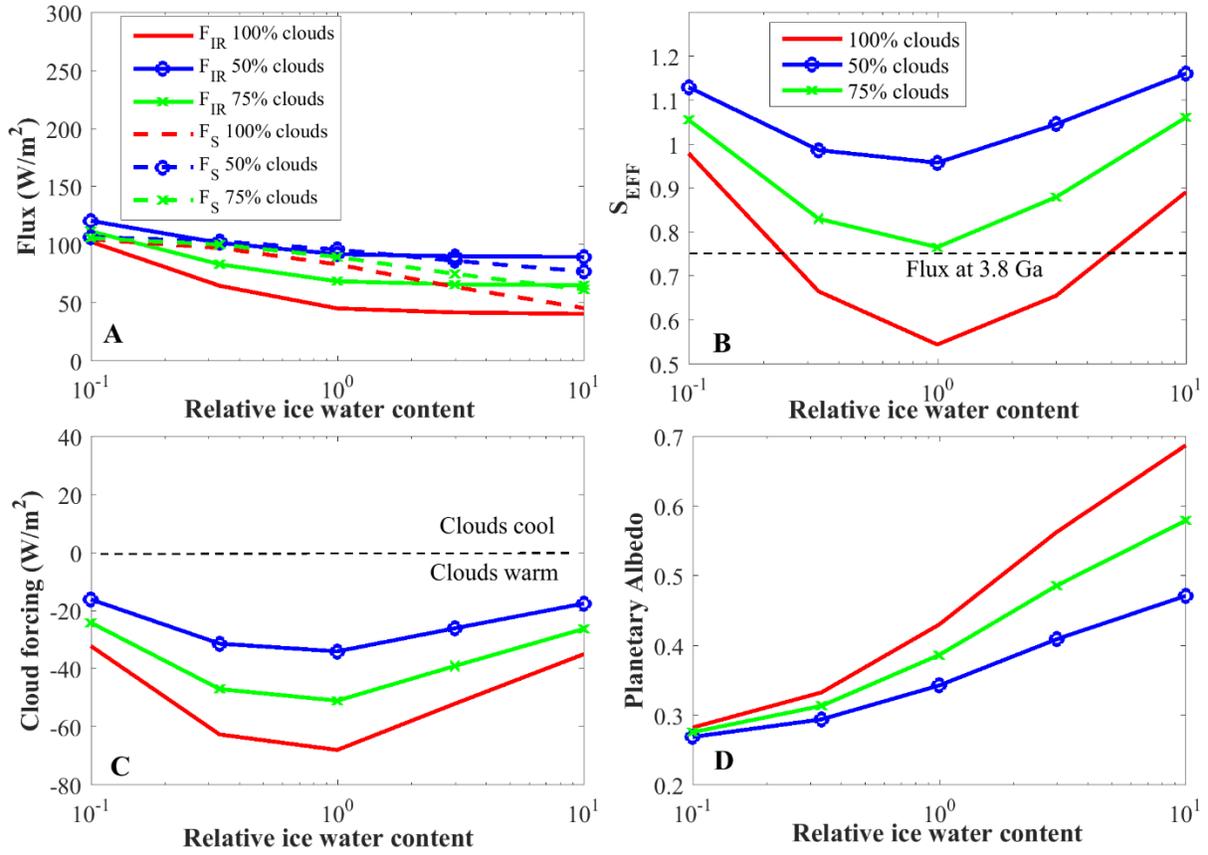

**Fig. 8:** Same as in Fig. 3 but using 5-km thick cirrus clouds composed of 100-micron cloud particles in the 0.5 bar atmosphere from Fig. 1

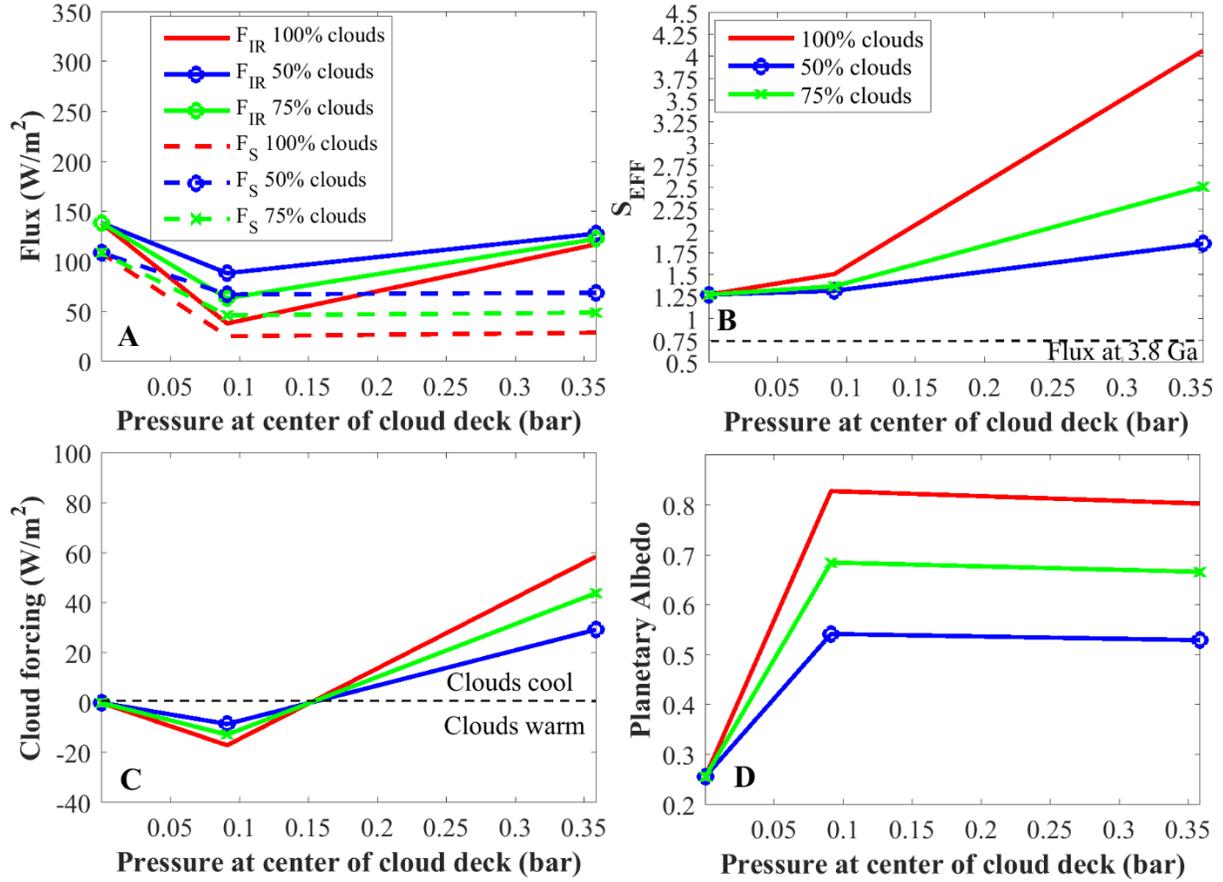

**Fig. 9:** Same as in Fig. 5 but using 10-km thick cirrus clouds composed of 10-micron cloud particles in the 0.5 bar atmosphere from Fig. 1

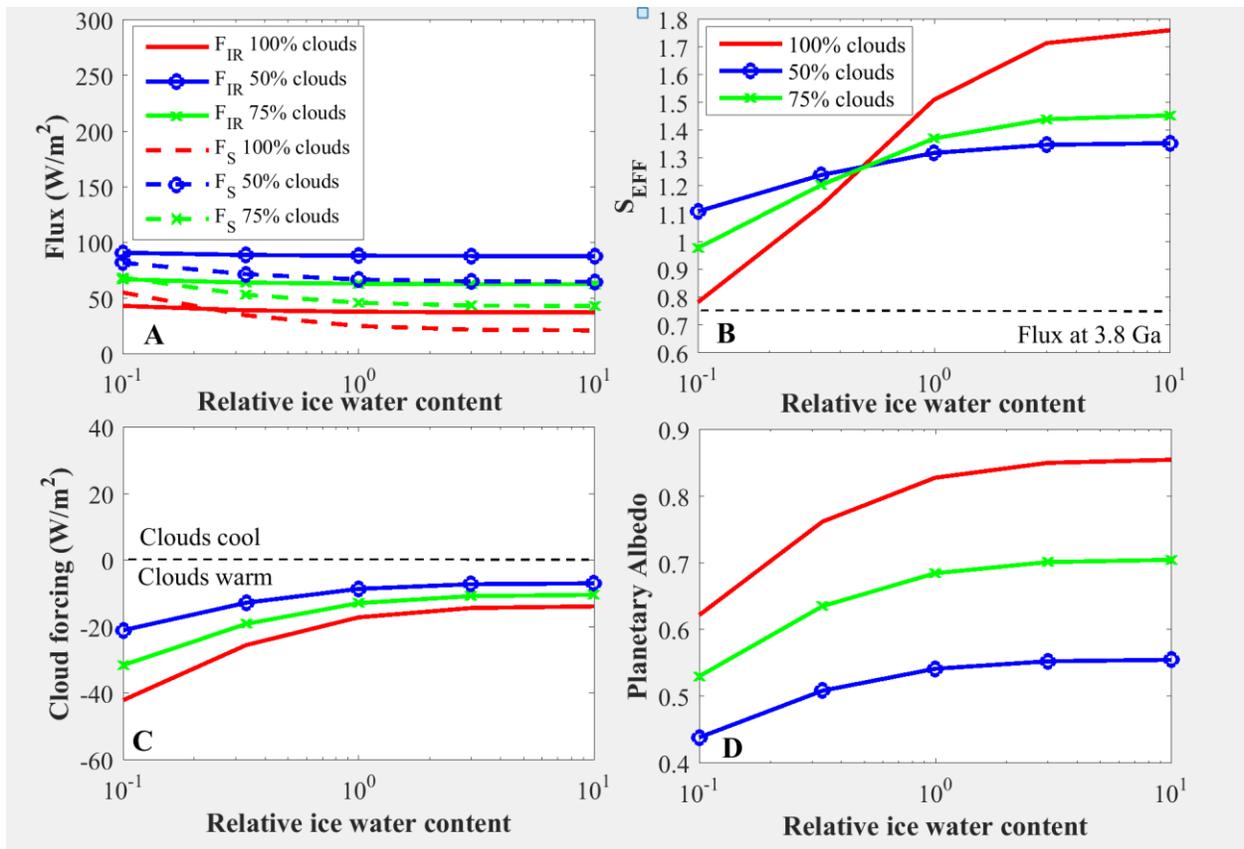

**Fig. 10:** Same as in Fig. 3 but using 10-km thick cirrus clouds composed of 10-micron cloud particles in the 0.5 bar atmosphere from Fig. 1

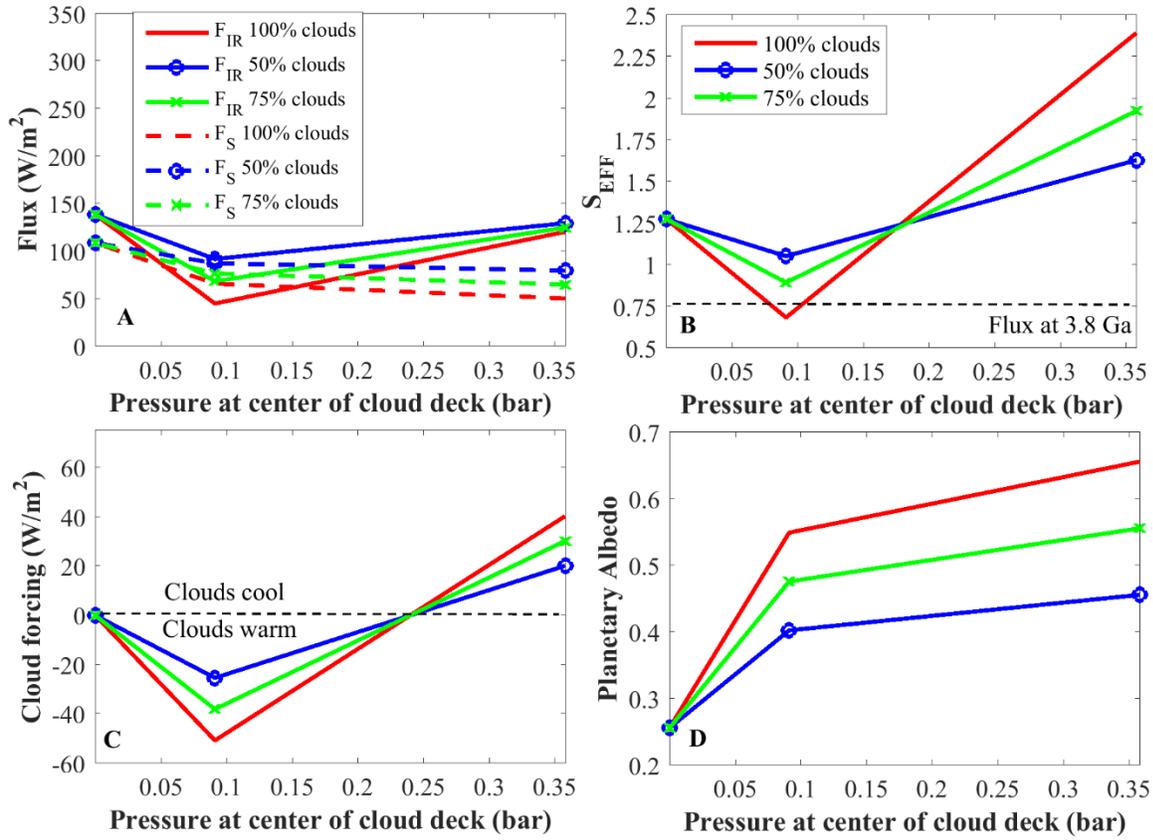

**Fig. 11:** Same as in Fig. 5 but using 10-km thick cirrus clouds composed of 100-micron cloud particles in the 0.5 bar atmosphere from Fig. 1

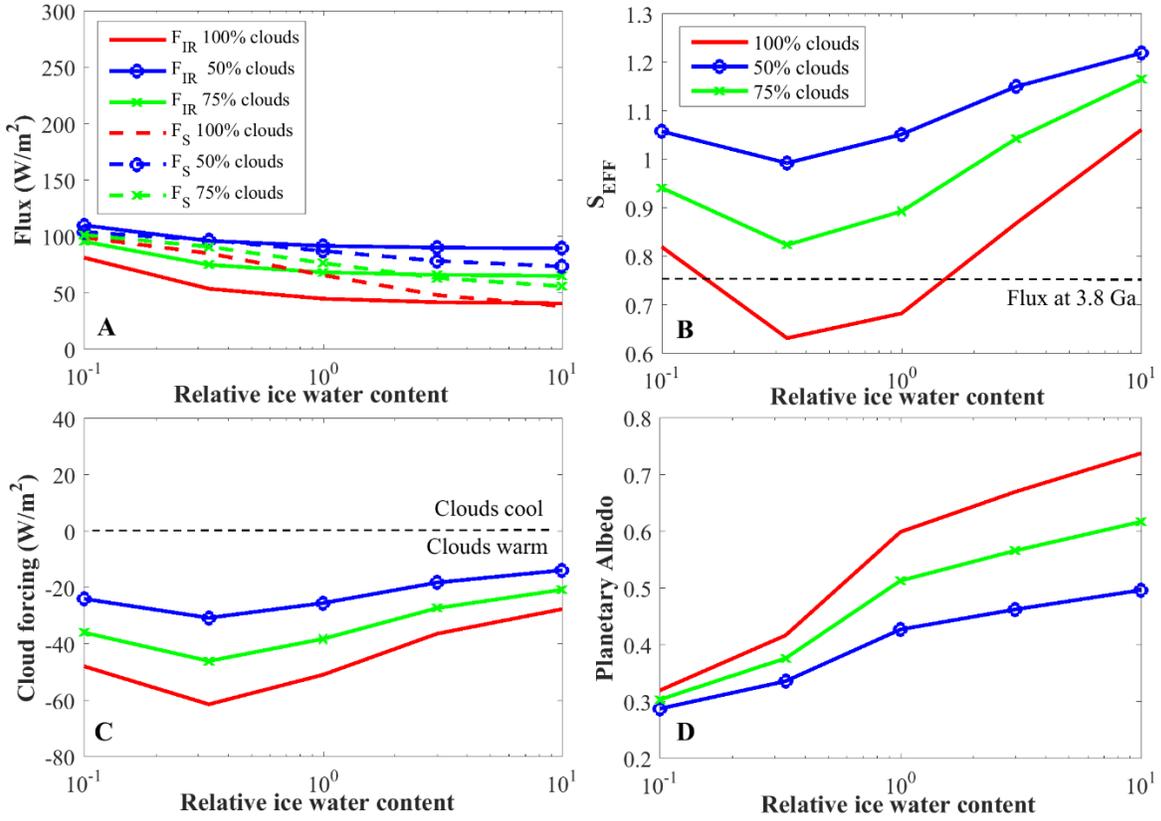

**Fig. 12:** Same as in Fig. 3 but using 10-km thick cirrus clouds composed of 100-micron cloud particles in the 0.5 bar atmosphere from Fig. 1

**Table Va: Sensitivity study for a 0.5 bar $CO_2$ present Mars atmosphere at P = $7.03 \times 10^{-2}$ bar with 5-km cirrus cloud decks and 10-micron particles with 100% cloud cover**

| IWC  | FIR   | FS     | SEFF   | PALB   | TAU(0.55 microns) | TAU(10 microns) |
|------|-------|--------|--------|--------|-------------------|-----------------|
| x1   | 38    | 33.19  | 1.145  | 0.773  | 50.06             | 58.56           |
| x3   | 37.55 | 24.58  | 1.528  | 0.8319 | 150.19            | 175.67          |
| x10  | 37.38 | 21.624 | 1.728  | 0.8521 | 500.63            | 585.55          |
| x1/3 | 39.31 | 50.96  | 0.7715 | 0.6515 | 16.52             | 19.32           |
| x1/10| 44.93 | 75.54  | 0.595  | 0.4833 | 5.006             | 5.856           |

**Table Vb: Sensitivity study for a 0.5 bar $CO_2$ present Mars atmosphere at P = $7.03 \times 10^{-2}$ bar with 5-km cirrus cloud decks and 100-micron particles with 100% cloud cover**

| IWC  | FIR   | FS    | SEFF   | PALB   | TAU(0.55 microns) | TAU(10 microns) |
|------|-------|-------|--------|--------|-------------------|-----------------|
| x1   | 45.35 | 83.29 | 0.5445 | 0.4303 | 5.06              | 5.305           |
| x3   | 41.86 | 63.86 | 0.6555 | 0.5632 | 16.18             | 15.915          |
| x10  | 40.74 | 45.69 | 0.8917 | 0.6875 | 50.6              | 53.05           |
| x1/3 | 64.86 | 97.47 | 0.6654 | 0.333  | 1.69              | 1.768           |
| x1/10| 102.8 | 104.9 | 0.98   | 0.2826 | 0.506             | 0.5305          |

**Table VIa: 5-km cirrus cloud decks and associated properties for 10 μm cirrus particles for the 0.5 bar $CO_2$ atmosphere at 100% cloud cover** [a]

| P | ALT | T (K) | FIR | FS | SEFF | PALB | IWC (g/m$^3$) | Tau(0.55 microns) | Tau(10 microns) |
|---|---|---|---|---|---|---|---|---|---|
| 0.428 | 2.38 | 267.14 | 131.62 | 36.31 | 3.62 | 0.7516 | 0.3765 | 309.1 | 361.5 |
| 0.288 | 7.76 | 250.95 | 115.66 | 29.8 | 3.88 | 0.7962 | 0.2531 | 219.98 | 257.3 |
| 0.189 | 13.03 | 230 | 89.74 | 27.53 | 3.259 | 0.8117 | 0.1665 | 135.37 | 158.33 |
| 0.117 | 18.5 | 204.27 | 59.72 | 28.11 | 2.125 | 0.8077 | 0.103 | 92.64 | 108.36 |
| 7.03(-2) | 23.57 | 178.94 | 38 | 33.19 | 1.145 | 0.7730 | 6.19(-2) | 50.06 | 58.55 |
| 5(-5) | 82.64 | 155 | 138.67 | 108.8 | 1.275 | 0.2556 | - | - | - |

**Table VIb: 5-km cirrus cloud decks and associated properties for 100 μm cirrus particles for the 0.5 bar $CO_2$ atmosphere at 100% cloud cover** [a]

| P | ALT | T (K) | FIR | FS | SEFF | PALB | IWC (g/m$^3$) | Tau(0.55 microns) | Tau(10 microns) |
|---|---|---|---|---|---|---|---|---|---|
| 0.428 | 2.38 | 267.14 | 132.52 | 60.2 | 2.21 | 0.5882 | 0.3764 | 30.91 | 32.75 |
| 0.288 | 7.76 | 250.95 | 119.25 | 60.74 | 1.963 | 0.5845 | 0.2531 | 22 | 23.31 |
| 0.189 | 13.03 | 230 | 96.98 | 66.84 | 1.451 | 0.5428 | 0.167 | 13.54 | 14.343 |
| 0.117 | 18.5 | 204.27 | 67.61 | 72.96 | 0.927 | 0.501 | 0.103 | 9.264 | 9.816 |
| 7.03(-2) | 23.57 | 178.94 | 45.35 | 83.29 | 0.544 | 0.4303 | 6.19(-2) | 5.06 | 5.305 |
| 5(-5) | 82.64 | 155 | 138.67 | 108.8 | 1.275 | 0.2556 | - | - | - |

[a] Values are listed at the midpoint of the cloud deck. Read 1.0(-5) as 1x10$^{-5}$.

**Table VIIa: 10-km cirrus cloud decks and associated properties for 10 µm cirrus particles for the 0.5 bar $CO_2$ atmosphere at 100% cloud cover** [a]

| P | ALT | T (K) | FIR | FS | SEFF | PALB | IWC (g/m$^3$) | Tau(0.55 microns) | Tau(10 microns) |
|---|---|---|---|---|---|---|---|---|---|
| 0.358 | 4.815 | 260.4 | 117.16 | 28.8 | 4.069 | 0.803 | 0.3155 | 517.35 | 605.11 |
| 9.13(-2) | 21.05 | 191.66 | 38.0 | 25.16 | 1.511 | 0.8279 | 8.04(-2) | 134.97 | 157.87 |
| 5(-5) | 82.64 | 155 | 138.67 | 108.8 | 1.274 | 0.2556 | - | - | - |

**Table VIIb: 10-km cirrus cloud decks and associated properties for 100 µm cirrus particles for the 0.5 bar $CO_2$ atmosphere at 100% cloud cover** [a]

| P | ALT | T (K) | FIR | FS | SEFF | PALB | IWC (g/m$^3$) | Tau(0.55 microns) | Tau(10 microns) |
|---|---|---|---|---|---|---|---|---|---|
| 0.358 | 4.815 | 260.4 | 120.4 | 50.33 | 2.394 | 0.6559 | 0.316 | 51.73 | 54.82 |
| 9.13(-2) | 21.05 | 191.66 | 44.98 | 65.89 | 0.6827 | 0.5493 | 8.04(-2) | 13.5 | 14.3 |
| 5(-5) | 82.64 | 155 | 138.67 | 108.8 | 1.274 | 0.2556 | - | - | - |

[a] Values are listed at the midpoint of the cloud deck. Read 1.0(-5) as 1x10$^{-5}$.

**Table VIII   Warm mean surface temperature values for the various cirrus cloud scenarios assuming 100% cloud coverage in the 0.5 bar early Mars ($S/S_o = 0.75$) atmosphere**
(*Indicates clouds did not produce mean surface temperatures above 273 K for these cases)

| | 1-km | | 5-km | | 10-km | |
|---|---|---|---|---|---|---|
| IWC | 10 microns(K) | 100 microns(K) | 10 microns(K) | 100 microns(K) | 10 microns(K) | 100 microns(K) |
| x 1/10 | * | * | 281 | * | * | * |
| X 1/3 | 284 | * | 273 | 279 | * | 279 |
| X 1 | 280 | * | * | 282 | * | 276 |
| X 3 | * | 285 | * | 277 | * | * |
| X 10 | * | 282 | * | * | * | * |

# 4. DISCUSSION

Our results suggest that the cirrus cloud warming mechanism of Urata and Toon (2013) could conceivably work, but it would require just the right circumstances. We find that mean surface temperatures above the freezing point of water could be maintained in our baseline 0.5 bar $CO_2$ atmosphere if ~ 1 km thick cirrus clouds decks composed of larger (e.g. 100 micron) particles are located at just the right atmospheric level (~0.06 bar) and if the cloud cover fraction is at least ~75%. Warm solutions with thicker clouds require nearly global cloud coverage. In reality, global cirrus cloud cover fractions above even 50*%* may be difficult to sustain on a planet with abundant precipitation. On Earth, global *total* cloud cover (which includes cirrus and all other cloud types) averages about 50% percent because water vapor condenses mostly in updrafts, while downdrafts are generally cloud-free. Thus, Earth's efficient rain-forming mechanism explains the partial cloud cover on our planet (Bartlett and Hunt, 1972). In other words, very high cloud cover fractions are difficult to achieve because subsidence regions will always be unsaturated. Moreover, out of this ~50% terrestrial global cloud cover, only half of this, or ~25-30% of the planetary surface, is covered in cirrus clouds (e.g., Wylie et al. 1994). Recent 3-D GCM simulations also compute that total cloud cover in a $CO_2$-dominated early Mars atmosphere would also be ~50% (Forget et al., 2013). Thus, it seems likely that cirrus cloud coverage on early Mars should also have been well under 50%. For the cirrus cloud mechanism to work, early Mars would have needed to exhibit cirrus cloud fractions that are much higher than those observed on Earth.

An additional reason for being skeptical about the cirrus cloud warming mechanism is the following. As mentioned in the Introduction, Urata and Toon (2013) argued that the auto-conversion rates of ice to snow cloud particles on early Mars could be up to a factor of 100 slower than those predicted for Earth. On Earth, cirrus cloud heights vary greatly from temperate to tropical regions, with altitudes ranging from ~ 2-18 km (e.g. Comstock and Jakob, 2004; Mace et al., 2005). Our 1-D model does not distinguish tropical from temperate regions, and our highest cirrus clouds for Earth are located somewhat lower than 10 km. In our Mars runs, cirrus cloud heights for our 0.5 and 3 bar atmospheres extend up to ~25 km and 18 km, respectively (Tables I–III). Applying Stokes' Law, we find that the cloud particle fall time for Earth from 10 km is nearly 3.5 hours. For Mars, these durations increase to ~18 hours and 13 hours for the 0.5 and 3 bar atmospheres, respectively. Should the cloud base on Mars also be located at 10 km, the corresponding fall time is just over 7 hours, or more than a factor of two greater than for the Earth, owing to the reduced gravity on Mars. Thus, cloud residence times in an early Mars atmosphere with cirrus clouds appear to be about 4-5 times longer than those for Earth which suggests that the 100-fold decrease in auto-conversion efficiency employed by Urata and Toon (2013) is a factor of 20 higher than is physically reasonable.

Urata and Toon (2013) further contend that relatively lower numbers of ice nuclei expected in the martian atmosphere lead to lower rates of coalescence and collection, resulting in lower precipitation rates. However, lower numbers of ice nuclei also imply bigger particles, which should lead to higher sedimentation rates, and thus to a decrease in cloud IWC (Prabhaka et al., 1993; Boehm et al., 1999) and a corresponding decrease in cloud radiative forcing. As fall speeds are proportional to $r^2$, according to Stokes' Law, 100-micron particles should fall out of the clouds ~100 times faster than equivalent 10-micron particles, leading to lower cloud optical depths and an *increase* in auto-conversion efficiency. This last assertion could be checked with more detailed cloud and microphysical models. Nevertheless, for all of these reasons, we believe that the low auto-conversion rates used by Urata and Toon are not physically reasonable.

A final question is whether we are being too stringent in requiring a mean surface temperature of 273 K. Could the fluvial features have formed under a somewhat colder climate and, if so, could cirrus clouds have provided the required warming? We tested this hypothesis by repeating the calculations shown in Fig. 7 for a mean surface temperature of 265 K (8 degrees lower than the surface temperature assumed in most of our calculations). $S_{eff}$ values required to support these cooler atmospheres are slightly lower than those found in Fig. 7: In Fig. 7, although 75% cloud cover is not quite enough to support the warm climate it suffices at the lower surface temperature ($S_{eff}$ = ~0.68). At 100% cloud cover, $S_{eff}$ is equal to 0.48 whereas it is 0.54 for Fig. 7. As with all of our simulations, 50% cloud cover is insufficient to support a warm climate, yielding an $S_{eff}$ of 0.87, which is well over the target value of 0.75 (In Fig. 7 $S_{eff}$ is 0.96). However, even this final simulation may be too optimistic because 265 K is probably well below the mean surface temperature that would allow large amounts of flowing water on the martian surface. Global mean surface temperatures during the Last Glacial Maximum on Earth were only about 3–7 K lower than today (Deimling et al., 2006; Annan and Hargreaves, 2013). If temperatures drop much below this, ice albedo feedback takes over, and global glaciation ensues, resulting in a mean surface temperature decrease of many tens of degrees (Crowley et al., 2001). An early Mars with smaller oceans than Earth might be less susceptible to ice-albedo feedback, and this hypothesis could be pursued with 3-D climate models.

**5. CONCLUSION**
Using a single-column radiative-convective climate model, we showed that cirrus clouds under the right circumstances (0.5 bar atmosphere, 1-km thin cirrus clouds) could have provided the necessary greenhouse warming to support a warm early Mars at cloud fractions as low as ~75%. This conclusion is in agreement that that of Urata and Toon (2013). However, this mechanism is probably not realistic because the efficacy of this mechanism decreases with thicker clouds, higher atmospheric $CO_2$ pressures, and smaller particle sizes. Moreover, actual cirrus cloud coverage on a warm early Mars would have likely been well below 50%. In addition, the very low cloud particle-to-precipitation auto-conversion rates assumed by Urata and Toon are probably not physically reasonable, and so may overestimate the effectiveness of the cirrus cloud warming mechanism. Larger cloud particles produce stronger radiative forcing, but they would have correspondingly shorter residence times and should thus decrease the net efficiency of cloud warming. Thus, other explanations are likely required to produce warm climates on early Mars.

**Acknowledgements:**
We would like to thank Richard Urata and Brian Toon for kindly providing the cloud optical data. RR acknowledges support by the Simons Foundation (SCOL 290357, LK), Carl Sagan Institute, and the Cornell Center of Astrophysics and Planetary Science. JK acknowledges support from NASA's Emerging Worlds, Habitable Worlds, and the Astrobiology Institute.